\documentclass{article}

\usepackage{calc}
\usepackage{amsmath}
\usepackage{amsthm}
\usepackage{amsfonts}
\usepackage{cite}
\usepackage{algorithm}
\usepackage{fullpage}
\usepackage[noend]{algorithmic}

\newtheorem{theorem}{Theorem}

\renewcommand{\Pr}{\mathbf{Pr}}
\newcommand{\E}{\mathbf{E}}

\def\I{{\cal I}}
\def\D{{\cal D}}

\begin{document}

\title{An Efficient Rigorous Approach for \\
Identifying Statistically Significant Frequent Itemsets\thanks{A preliminary version of this work was presented in ACM PODS
2009.}}
\author{
Adam Kirsch\thanks{Harvard School of Engineering and Applied Sciences, Cambridge, MA, USA. Email: \texttt{kirsch@eecs.harvard.edu}. Supported in part by NSF Grant CNS-0721491 and a grant from
Cisco Systems Inc.}
\\
\and
\vspace{0.1cm}
Michael Mitzenmacher\thanks{Harvard School of Engineering and Applied Sciences, Cambridge, MA, USA. Email: \texttt{michaelm@eecs.harvard.edu}. Supported in part by NSF Grant CNS-0721491 and grants from Cisco
Systems Inc., Yahoo!, and Google.}
\\
\and
\vspace{0.1cm}
Andrea Pietracaprina\thanks{Department of Information Engineering,
  University of Padova, Italy. Email:
  \texttt{andrea.pietracaprina@unipd.it}. Supported in part by the EC/IST Project 15964 AEOLUS.}
\and
Geppino Pucci\thanks{Department of Information Engineering, University
  of Padova, Italy. Email: \texttt{geppino.pucci@unipd.it}. Supported in part by the EC/IST Project 15964 AEOLUS.}
\and
Eli Upfal\thanks{Computer Science Department, Brown University, Providence, RI, USA. Email: \texttt{eli@cs.brown.edu}. Supported in part by NSF awards IIS-0325838 and DMI-0600384, ONR Award
N000140610607, and EC/IST Project 15964 AEOLUS.}
\and
Fabio Vandin\thanks{Department of Information Engineering, University
  of Padova, Italy. Email: \texttt{vandinfa@dei.unipd.it}. Supported in part by the EC/IST Project 15964 AEOLUS.
}
}

\maketitle

\begin{abstract}
As advances in technology allow for the collection, storage, and
analysis of vast amounts of data, the task of screening and assessing
the significance of discovered patterns is becoming a major challenge
in data mining applications.  In this work, we address significance in
the context of frequent itemset mining. Specifically, we develop a
novel methodology to identify a meaningful support threshold $s^*$ for
a dataset, such that the number of itemsets with support at least $s^*$
represents a substantial deviation from what would be expected in a random
dataset with the same number of transactions and the same individual
item frequencies. These itemsets can then be flagged as statistically
significant with a small false discovery rate.
We present extensive experimental results to substantiate the
effectiveness of our methodology.
\end{abstract}

\section{Introduction}

The discovery of frequent itemsets in transactional datasets is
 a fundamental primitive that arises in the mining of
association rules and in many other scenarios
\cite{HanK01,TanSK06}. In its original formulation, the problem
requires that given a dataset $\D$ of transactions over a set of items
$\I$, and a support threshold $s$, all itemsets $X \subseteq \I$ with
support at least $s$ in $\D$ (i.e., contained in at least $s$ transactions) be
returned. These high-support itemsets are referred to as
\emph{frequent itemsets}.

Since the pioneering paper by Agrawal et al.~\cite{AgrawalIS93}, a
vast literature has flourished, addressing variants of the problem,
studying foundational issues, and presenting novel algorithmic
strategies or clever implementations of known strategies (see, e.g.,
\cite{Fimi03,Fimi04}), but many problems remain open
\cite{HanCXY07}. In particular, assessing the significance of the
discovered itemsets, or equivalently, flagging statistically
significant discoveries with a limited number of false positive
outcomes, is still poorly understood and remains one of the most
challenging problems in this area.

The classical framework requires that the user decide what is
significant by specifying the support threshold $s$.  Unless specific
domain knowledge is available, the choice of such a threshold is often
arbitrary \cite{HanK01,TanSK06} and may lead to a large number of
spurious discoveries (false positives) that would undermine the
success of subsequent analysis.

In this paper, we develop a rigorous and efficient novel approach for
identifying frequent itemsets featuring both a global and a pointwise
guarantee on their statistical significance.  Specifically, we flag as
significant a population of itemsets extracted with respect to a
certain threshold, if some global characteristics of the population
deviate considerably from what would be expected if the dataset were
generated randomly with no correlations between items. Also, we make sure that
a large fraction of the itemsets belonging to the returned population
are individually significant by enforcing a small False Discovery Rate
(FDR)~\cite{BenjaminiH95} for the population.

\subsection{The model} \label{model}
As mentioned above, the significance of a discovery in our framework
is assessed based on its deviation from what would be expected in a
random dataset in which individual items are placed in transactions
independently.

Formally, let $\D$ be a dataset of $t$ transactions on a set $\I$ of
$n$ items, where each transaction is a subset of $\I$. Let $n(i)$ be
the number of transactions that contain item $i$ and let $f_i=n(i)/t$
be the {\emph{frequency}} of item $i$ in the dataset.  The
{\emph{support}} of an itemset $X \subseteq \I$ is defined as the
number of transactions that contain $X$.
Following~\cite{SilversteinBM98}, the dataset $\D$ is associated with
a probability space of datasets, all featuring the same number of
transactions $t$ on the same set of items $\I$ as $\D$, and in which
item $i$ is included in any given transaction with probability $f_i$,
independently of all other items and all other transactions. A similar
model is used in \cite{PurdomVGG04} and \cite{SayrafiVGP05} to
evaluate the running time of algorithms for frequent itemsets mining.
For a fixed integer $k \geq 1$, among all possible $n\choose k$
itemsets of size $k$ (\emph{$k$-itemsets}) we are interested in
identifying statistically significant ones, that is, those $k$-itemsets
whose supports are significantly higher, in a statistical sense, than
their expected supports in a dataset drawn at random from the
aforementioned probability space.

An alternative probability space of datasets, proposed in
\cite{GionisMMT06}, considers all arrangements of $n$ items into $m$
transactions which match the exact item frequencies and transaction
lengths as $\D$. Conceivably, the technique of this paper could be
adapted to this latter model as well.

\subsection{Multi-hypothesis testing} \label{sec:multi-hp}

In a simple statistical test, a null hypothesis $H_0$ is tested against
an alternative hypothesis $H_1$.  A test consists of a rejection
(critical) region $C$ such that, if the statistic (outcome) of the
experiment is in $C$, then the null hypothesis is rejected, and otherwise the
null hypothesis is not rejected. The {\it significance level} of a
test, $\alpha=\Pr$(Type I error), is the probability of rejecting $H_0$
when it is true (false positive).  The {\it power} of the test,
$1-\Pr$(Type II error), is the probability of correctly rejecting the
null hypothesis.  The {\it $p$-value} of a test is the probability of
obtaining an outcome at least as extreme as the one that was actually
observed, under the assumption that $H_0$ is true.

In a multi-hypothesis statistical test, the outcome of an experiment
is used to test simultaneously a number of hypotheses. For example, in
the context of frequent itemsets, if we seek significant $k$-itemsets,
we are in principle testing $n \choose k$ null hypotheses simultaneously,
where each null hypothesis corresponds to the support of a given
itemset not being statistically significant.  In the context of
multi-hypothesis testing, the significance level cannot be assessed by
considering each individual hypothesis in isolation.  To demonstrate
the importance of correcting for multiplicity of hypotheses, consider
a simple real dataset of 1,000,000 transactions over 1,000 items, each
with frequency 1/1,000.  Assume that we observed that a pair of items
$(i,j)$ appears in at least 7 transactions.  Is the support of this
pair statistically significant? To evaluate the significance of this
discovery we consider a random dataset where each item is included in
each transaction with probability 1/1,000, independent of all items.
The probability that the pair $(i,j)$ is included in a given
transaction is 1/1,000,000, thus the expected number of transactions
that include this pair is 1.  A simple calculation shows that the
probability that $(i,j)$ appears in at least 7 transactions is about
0.0001.  Thus, it seems that the support of $(i,j)$ in the real
dataset is statistically significant.  However, each of the 499,500
pairs of items has probability 0.0001 to appear in at least 7
transactions in the random dataset. Thus, even under the assumption
that items are placed independently in transactions, the expected
number of pairs with support at least 7 is about 50. If there were
only about 50 pairs with support at least 7, returning the pair
$(i,j)$ as a statistically significant itemset would likely be a false
discovery since its frequency would be better explained by random
fluctuations in observed data.  On the other hand, assume that the
real dataset contains 300 disjoint pairs each with support at least 7.
By the Chernoff bound \cite{mu05}, the probability of that event in
the random dataset is less than $2^{-300}$. Thus, it is very likely
that the support of most of these pairs would be statistically
significant. A discovery process that does not return these pairs will
result in a large number of false negative errors.  Our goal is to
design a rigorous methodology which is able to distinguish between
these two scenarios.

A natural generalization of the significance level to multi-hypothesis
testing is the {\it Family Wise Error Rate (FWER)}, which is the
probability of incurring at least one Type I error in any of the
individual tests.  If we have $m$ simultaneous tests and we want to
bound the FWER by $\alpha$, then the Bonferroni method tests each null
hypothesis with significance level $\alpha/m$. While controlling the
FWER, this method is too conservative in that the power of the test is
too low, giving many false negatives. There are a number of techniques
that improve on the Bonferroni method, but for large numbers of
hypotheses all of these techniques lead to tests with low power (see
\cite{Dudoit03} for a good review).

The {\it False Discovery Rate (FDR)} was suggested by Benjamini and
Hochberg~\cite{BenjaminiH95} as an alternative, less conservative
approach to control errors in multiple tests. Let $V$ be the number of
Type I errors in the individual tests, and let $R$ be the total number
of null hypotheses rejected by the multiple test. Then we define FDR
to be the expected ratio of erroneous rejections among all rejections,
namely FDR $= E[V/R]$, with $V/R=0$ when $R=0$. Designing a
statistical test that controls for FDR is not simple, since the FDR is
a function of two random variables that depend both on the set of null
hypotheses and the set of alternative hypotheses.  Building on the
work of \cite{BenjaminiH95}, Benjamini and Yekutieli~\cite{BY01}
developed a general technique for controlling the FDR in any
multi-hypothesis test (see Theorem~\ref{by-test} in
Section~\ref{sec:chernoff}).

\subsection{Our Results}\label{sec:ourresults}

We address the classical problem of mining frequent itemsets with
respect to a certain minimum support threshold, and provide a rigorous
methodology to establish a threshold that guarantees, in a statistical
sense, that the returned family of frequent itemsets contains
significant ones with a limited FDR. Our methodology crucially relies
on the following Poisson approximation result, which is the main
theoretical contribution of the paper.

Consider a dataset $\D$ of $t$ transactions on a set $\I$ of $n$ items
and let $\hat{\D}$ be a corresponding random dataset according to
the random model described in Section~\ref{model}. Let $Q_{k,s}$
be the observed number of $k$-itemsets with support at least $s$
in $\D$, and let $\hat{Q}_{k,s}$ be the corresponding
random variable for $\hat{\cal D}$. We show that there exists a
minimum support value $s_{\min}$ (which depends on the parameters of
$\D$ and on $k$), such that for all $s \ge s_{\min}$ the distribution of
$\hat{Q}_{k,s}$ is well approximated by a Poisson distribution.  Our
result is based on a novel application of the Chen-Stein Poisson
approximation method \cite{ArratiaGG90}.  

The minimum support $s_{\min}$ provides the grounds to devise a
rigorous method for establishing a support threshold for mining
significant itemsets, both reducing the overall complexity and
improving the accuracy of the discovery process.  Specifically, for a
fixed itemset size $k$, we test a small number of support thresholds
$s \geq s_{\min}$, and, for each such threshold, we measure the
$p$-value corresponding to the null hypothesis $H_0$ that the observed
value $Q_{k,s}$ comes from a Poisson distribution of suitable
expectation.  From the tests we can determine a threshold $s^*$ such
that, with user-defined significance level $\alpha$, the number of
$k$-itemsets with support at least $s^*$ is not sampled from a Poisson
distribution and is therefore statistically significant. Observe that
the statistical significance of the number of itemsets with support at
least $s^*$ does not imply necessarily that each of the itemsets is
significant. However, our test is also able to guarantee a
user-defined upper bound $\beta$ on the FDR among all discoveries.  We
remark that our approach works for any fixed itemset size $k$, unlike
traditional frequent itemset mining, where itemsets of all sizes are
extracted for a given threshold.

To grasp the intuition behind the above approach, recall that a
Poisson distribution models the number of occurrences among a large
set of possible events, where the probability of each event is
small. In the context of frequent itemset mining, the Poisson
approximation holds when the probability that an individual itemset
has support at least $s_{\min}$ in $\hat{\D}$ is small, and thus the
existence of such an event in $\D$ is likely to be statistically
significant.  We stress that our technique discovers statistically
significant itemsets among those of relatively high support. In fact,
if the expected supports of individual itemsets vary in a large range,
there may exist itemsets with very low expected supports in $\hat{\D}$
which may have statistically significant supports in $\D$. These
itemsets would not be discovered by our strategy.  However, any mining
strategy aiming at discovering significant, low-support itemsets is
likely to incur high costs due to the large (possibly exponential)
number of candidates to be examined, although only a few of them would turn
out to be significant.

We validate our theoretical results by mining significant frequent
itemsets from a number of real datasets that are standard benchmarks
in this field.  Also, we compare the performance of our methodology to
a standard multi-hypothesis approach based on \cite{BY01}, and provide
evidence that the latter often returns fewer significant itemsets,
which indicates that our method has considerably higher power.

\subsection{Related Work}

A number of works have explored various notions of significant
itemsets and have proposed methods for their discovery. Below, we review those
most relevant to our approach and refer the reader to
\cite[Section~3]{HanCXY07} for further references. 
Aggarwal and Yu \cite{AggarwalY98} relate the significance of an itemset
$X$ to the
quantity $((1-v(X))/(1-\E[v(X)])) \cdot (\E[v(X)]/v(X))$, where $v(X)$
represents the fraction of transactions containing some but not all of
the items of $X$, and $\E[v(X)]$ represents the expectation of $v(X)$
in a random dataset where items occur in transactions
independently. This ratio provides an empirical measure of the
correlation among the items of $X$ that, according
to the authors, is more effective than absolute support. In
\cite{SrikantA96-1,DuMouchel99,DuMouchelP01}, the significance of an
itemset is measured as the ratio $R$ between its actual support and
its expected support in a random dataset. In order to make this
measure more accurate for small supports,
\cite{DuMouchel99,DuMouchelP01} propose smoothing the ratio $R$ using
an empirical Bayesian approach. Bayesian analysis is also employed in
\cite{SilberschatzT96} to derive subjective measures of significance
of patterns (e.g., itemsets) based on how strongly they ``shake'' a
system of established beliefs. In \cite{JaroszewiczS05}, the 
significance of an itemset is defined as the absolute difference 
between the support of the itemset in the dataset, and the estimate
of this support made from a Bayesian network with parameters derived from the dataset.

A statistical approach for identifying significant itemsets is
presented in \cite{SilversteinBM98}, where the measure of interest for
an itemset is defined as the degree of dependence among its
constituent items, which is assessed through a $\chi^2$ test.
Unfortunately, as reported in~\cite{DuMouchel99,DuMouchelP01}, there
are technical flaws in the applications of the statistical test in
\cite{SilversteinBM98}.  Nevertheless, this work
pioneered the quest for a rigorous framework for addressing the
discovery of significant itemsets.

A common drawback of the aforementioned works is that they assess the
significance of each itemset \emph{in isolation}, rather than taking
into account the \emph{global} characteristics of the dataset from
which they are extracted. As argued before, if the number of itemsets
considered by the analysis is large, even in a purely random dataset
some of them are likely to be flagged as significant if considered in
isolation.  A few works attempt at accounting for the global structure
of the dataset in the context of frequent itemset mining.  The authors
of \cite{GionisMMT06} propose an approach based on Markov chains to
generate a random dataset that has identical transaction lengths and
identical frequencies of the individual items as the given real
dataset.  The work suggests comparing the outcomes of a number of data
mining tasks, frequent itemset mining among the others, in the real
and the randomly generated datasets in order to assess whether the
real datasets embody any significant global structure.  However, such
an assessment is carried out in a purely qualitative fashion without
rigorous statistical grounding.

The problem of spurious discoveries in the mining of significant patterns
is studied in \cite{BoltonHA02}. The paper is concerned with the
discovery of significant pairs of items, where significance is
measured through the $p$-value, that is, the probability of occurrence
of the observed support in a random dataset. Significant pairs are
those whose $p$-values are below a certain threshold that can be
suitably chosen to bound the FWER, or to bound the FDR. The authors
compare the relative power of the two metrics through experimental results, 
but do not provide
methods to set a meaningful support threshold, which is the most
prominent feature of our approach.

Beyond frequent itemset mining, the issue of significance has also
been addressed in the realm of discovering association rules. In
\cite{HamalainenN08}, the authors provide a variation of the
well-known Apriori strategy for the efficient discovery of a subset
$\cal A$ of association rules with $p$-value below a given cutoff
value, while the results in \cite{MegiddoS98} provide the means of
evaluating the FDR in $\cal A$. The FDR metric is also employed in
\cite{ZhangPT04} for the discovery of significant
quantitative rules, a variation of association rules. None of these
works is able to establish support thresholds such that the returned
discoveries feature small FDR.

\subsection{Benchmark datasets}
\sloppy In order to validate the methodology, a number of experiments,
whose results are reported in Section~\ref{sec:experiments}, have been
performed on datasets which are standard benchmarks in the context of
frequent itemsets mining. The main characteristics of the datasets we
use are summarized in Table~\ref{tab:datasets}. A description of the
datasets can be found in the FIMI Repository ({\tt
  http://fimi.cs.helsinki.fi/data/}), where they are available for 
download.

\begin{table}
\begin{center}
\begin{tabular}{lcccc}
\hline
Dataset & $n$ & $\left[ f_{\min};f_{\max}\right]$ & $m$ & $t$\\
\hline
Retail & 16470 & $[$1.13e-05 ; 0.57$]$ &  10.3 & 88162 \\
Kosarak & 41270 & $[$1.01e-06 ; 0.61$]$ & 8.1 & 990002 \\
Bms1 & 497 & $[$1.68e-05 ; 0.06$]$ & 2.5 & 59602 \\
Bms2 & 3340 & $[$1.29e-05 ; 0.05$]$ & 5.6 & 77512 \\
Bmspos & 1657 & $[$1.94e-06 ; 0.60$]$ & 7.5 & 515597 \\
Pumsb$^*$ & 2088 & $[$2.04e-05 ; 0.79$]$ & 50.5 & 49046 \\
\hline
\end{tabular}
\end{center}
\caption{\boldmath Parameters of the benchmark
datasets: $n$ is the number of items; $[f_{\min},f_{\max}]$ is the
range of frequencies of the individual items; $m$ is the average
transaction length; and $t$ is the number of transactions.}
\label{tab:datasets}
\end{table}

\subsection{Organization of the Paper}\label{subsec:roadmap}

The rest of the paper is structured as follows.
Section~\ref{sec:poisson} presents the Poisson approximation
result for the random variable $\hat{Q}_{k,s}$.  The methodology
for establishing the support threshold $s^*$  is
presented in Section~\ref{sec:methodology}, and experimental
results are reported in Section~\ref{sec:experiments}. 
 Section~\ref{sec:conclusions} ends the paper with some concluding
remarks.

\section{Poisson Approximation Result}\label{sec:poisson}

The Chen-Stein method \cite{ArratiaGG90} is a powerful tool for
bounding the error in approximating probabilities associated with a
sequence of dependent events by a Poisson distribution.  To apply the
method to our case, we fix parameters $k$ and $s$, and define a
collection of ${n \choose k}$ Bernoulli random variables
$\{Z_X~|~X\subset {\cal I},~|X|=k\}$, such that $Z_X=1$ if the
$k$-itemset $X$ appears in at least $s$ transactions in the random
dataset $\hat{\cal D}$, and $Z_X=0$ otherwise.  Also, let $p_X =
\Pr(Z_X = 1)$. We are interested in the distribution of
$\hat{Q}_{k,s}=\sum_{X:|X|=k} Z_X$.

For each set $X$ we define the \emph{neighborhood set} of $X$,
\[
    I(X)=\{X'~|~X\cap X' \ne \emptyset, |X'| = |X|\}.
\]
If $Y \not\in I(X)$ then $Z_Y$ and $Z_X$ are independent. 
The following theorem is a straightforward adaptation
of \cite[Theorem 1]{ArratiaGG90} to our case.
\begin{theorem}\label{arratia}
Let $U$ be a Poisson random variable such that
$\E[U]=\E[\hat{Q}_{k,s}]=\lambda<\infty$. The variation distance
between the distributions ${\cal L}(\hat{Q}_{k,s})$ of $\hat{Q}_{k,s}$
and ${\cal L}(U)$ of $U$ is such that 
\begin{align*}
    \left\| {\cal L}(\hat{Q}_{k,s})-{\cal L}(U) \right\|
    &= \sup_A |\Pr(\hat{Q}_{k,s} \in A)-\Pr(U \in A)| \\
    &\leq b_1+b_2,
\end{align*}
where
\[
    b_1  =  \sum_{X:|X|=k} \sum_{Y\in I(X)} p_X p_Y
\]
and
\[
    b_2 = \sum_{X:|X|=k} \sum_{X\neq Y\in I(X)} \E[Z_X Z_Y].
\]
\end{theorem}

We can derive analytic bounds for $b_1$ and $b_2$ in many situations.
Specifically, suppose that we generate $t$ transactions in the
following way.  For each item $x$, we sample a random variable $R_x
\in [0,1]$ independently from some distribution $R$.  Conditioned on
the $R_x$'s, each item $x$ occurs independently in each transaction
with probability $R_x$.  In what follows, we provide specific bounds
for this situation that depend on the moment $\E[R^{2s}]$ of the
random variable $R$. 

As a warm-up, we first consider the specific case where each $R_x$ is
a fixed value $p = \gamma/n$ for some constant $\gamma$ for all $x$.
That is, each item appears in each transaction with a fixed
probability $p$, and the expected number of items per transaction is
constant.  The more general case follows the same approach, albeit
with a few more technical difficulties.

\begin{theorem} \label{thm:poissapproxsimple}
Consider an asymptotic regime where as $n \to \infty$, we have $k,s =
O(1)$ with $s \ge 2$, each item appears in each transaction with probability
$p = \gamma/n$ for some constant $\gamma$, 
and $t = O(n^c)$ for some positive constant $0 < c \leq (k-1)(1-1/s)$.
Let $U$ be a Poisson random variable such that
$\E[U]=\E[\hat{Q}_{k,s}]=\lambda<\infty$. 
Then the variation distance between the distributions ${\cal
L}(\hat{Q}_{k,s})$ of $\hat{Q}_{k,s}$ and
${\cal L}(U)$ of $U$ satisfies
\iffalse

\begin{align*}
    \left\| {\cal L}(\hat{Q}_{k,s})-{\cal L}(U) \right\|
    &= \sup_A |\Pr(\hat{Q}_{k,s} \in A)-\Pr(U \in A)|\\
    &\le b_1 + b_2,
\end{align*}
where
\begin{align*}
    b_1 &= \left( \binom{n}{k}^2 - \binom{n}{k}\binom{n-k}{k} \right)\\&\quad {} \times
            \left( \sum_{i=s}^t \binom{t}{i} \left(\frac{\gamma}{n}\right)^{ki} \left(1-\left(\frac{\gamma}{n}\right)^{k}\right)^{t-i} \right)^2\\
        &= \Theta(n^{2cs + 2k(1-s)-1})
\end{align*}
and
\begin{align*}
    b_2 &\le \sum_{g=1}^{k-1} \binom{n}{g ; k-g; k-g} \left(\frac{\gamma}{n}\right)^{2ks}\\&\qquad\qquad {} \times \sum_{i=0}^s \binom{t}{i;s-i;s-i} \left(\frac{n}{\gamma}\right)^{gi}\\
    &= O(n^{2k(1-s) + s(k-1+c)-k+1}).
\end{align*}
\fi
\[
    \left\| {\cal L}(\hat{Q}_{k,s})-{\cal L}(U) \right\| = O(1/n^{2s-2}).
\]
\end{theorem}
\begin{proof}

For a given set $X$ of $k$ items, let $p_{X,i}$ be the probability
that $S$ appears in exactly $i$ transactions, so that $p_X =
\sum_{i=s}^t p_{X,i}$ and
\[
    p_{X,i} = \binom{t}{i} \left(\frac{\gamma}{n}\right)^{ki} \left(1-\left(\frac{\gamma}{n}\right)^k\right)^{t-i}.
\]

Applying Theorem~\ref{arratia} gives
\[
    \left\|{\cal L}(\hat{Q}_{k,s})-{\cal L}(U)\right\| \leq b_1 + b_2
\]
where
\[
    b_1  =  \sum_{X:|X|=k} \sum_{Y\in I(S)} p_X p_Y
\]
and
\[
    b_2 = \sum_{X:|S|=k} \sum_{Y\neq X\in I(S)} \E[Z_X Z_Y].
\]

We now evaluate $b_1$ and $b_2$.  A direct calculation easily gives
the value for $b_1$ given in the statement of the theorem.  For the
asymptotic analysis, we write
\begin{align*}
    &\left(\binom{n}{k}^2 - \binom{n}{k}\binom{n-k}{k}\right)\\
    &=\binom{n}{k}^2\left(1 - \frac{\binom{n-k}{k}}{\binom{n}{k}}\right)\\
    &=\binom{n}{k}^2\left(1 - \prod_{i=0}^{k-1} \frac{n-k-i}{n-i}\right)\\
    &=\Theta(n^k)^2 \cdot \Theta(1/n)
    = \Theta(n^{2k-1})
\end{align*}
and
\begin{align*}
    p_{X,s} &= \binom{t}{s} \left(\frac{\gamma}{n}\right)^{ks} \left(1-\left(\frac{\gamma}{n}\right)^{k}\right)^{t-s}\\
    &=\Theta(t^s) \cdot \Theta(n^{-ks}) \cdot (1 + o(1))
    =\Theta\left(t^{s} n^{-ks}\right),
\end{align*}
where we have used the fact that $t = o(n^k)$ to obtain the
asymptotics for the third term.  Also, we note that for any $1 \le i
< t $
\[
    \frac{p_{X,i+1}}{p_{X,i}}
    = \frac{t-i}{i+1} \left(\frac{\gamma}{n}\right)^k\left(1-\left(\frac{\gamma}{n}\right)^k\right)^{-1}
\]
and so
\[
    \max_{i \in \{s,s+1,\ldots,t-1\}} \frac{p_{X,i+1}}{p_{X,i}} = O(tn^{-k}) = O(1/n).
\]

Using a geometric series, it follows that
\[
    p_X = \sum_{i=s}^t p_{X,i} = p_{X,s}(1 + o(1)) = \Theta\left(t^{s} n^{-ks}\right).
\]

Thus, we obtain
\begin{align*}
    b_1 &= \Theta(n^{2k-1})\cdot\Theta\left(t^{s} n^{-ks}\right)^2\\
    &= \Theta(t^{2s}n^{2k(1-s)-1})
    = \Theta(n^{2cs + 2k(1-s)-1}).
\end{align*}

We now turn our attention to $b_2$.  Consider sets $X \ne Y$ of $k$
items, let $g = |X \cap Y|$, and suppose that $g > 0$.  Then if $Z_X
Z_Y = 1$, there exist disjoint subsets $A,B,C \in \{1,\ldots,t\}$
such that $0 \le |A| \le s$, $|B| = |C| = s-|A|$, all of the
transactions in $A$ contain both $X$ and $Y$, all of the transactions
in $B$ contain $X$, and all of the transactions in $C$ contain $Y$.

Therefore,
\[
    \E[Z_X Z_Y]
    \le \sum_{i=0}^s \binom{t}{i; s-i; s-i} \left(\frac{\gamma}{n}\right)^{(2k-g)i + 2k(s-i)},
\]
where the notation $\binom{m}{x; y; z}$ is a shorthand for
$\binom{m}{x}\binom{m-x}{y}\binom{m-x-y}{z}$. 

It follows that
\begin{align*}
    b_2
    &\le \sum_{g=1}^{k-1} \binom{n}{g ; k-g; k-g} \\&\qquad {} \times \sum_{i=0}^s \binom{t}{i;s-i;s-i} \left(\frac{\gamma}{n}\right)^{(2k-g)i + 2k(s-i)}\\
    &= \sum_{g=1}^{k-1} \binom{n}{g ; k-g; k-g} \left(\frac{\gamma}{n}\right)^{2ks} \\&\qquad {} \times \sum_{i=0}^s \binom{t}{i;s-i;s-i} \left(\frac{n}{\gamma}\right)^{gi}\\    &= \sum_{g=1}^{k-1} \binom{n}{g ; k-g; k-g} \left(\frac{\gamma}{n}\right)^{2ks} \\&\qquad {} \times \sum_{i=0}^s \binom{t}{i;s-i;s-i} \left(\frac{n}{\gamma}\right)^{gi}\\
    &= \sum_{g=1}^{k-1} \Theta(n^{2k-g+2cs}) \left(\frac{\gamma}{n}\right)^{2ks} \sum_{i=0}^s n^{-ic} \left(\frac{n}{\gamma}\right)^{gi}\\
    &= \Theta(n^{2k(1-s) + 2cs})\sum_{g=1}^{k-1} n^{-g} \sum_{i=0}^s {\gamma}^{-gi} n^{(g-c)i}\\
    &= \Theta(n^{2k(1-s) + 2cs})\sum_{g=1}^{k-1} n^{-g} \begin{cases} \Theta(1) & g \le c\\ \Theta(n^{(g-c)s}) & g > c \end{cases}\\
    &= \Theta(n^{2k(1-s) + 2cs}) \cdot \Theta(n^{-(k-1) + (k-1-c)s})\\
    &= \Theta(n^{2k(1-s) + s(k-1+c)-k+1})
\end{align*}

Note that, in the summation where there are two cases depending on whether $g \leq c$ or $g > c$, 
we have used the assumption that $c \leq (k-1)(1-1/s)$ to ensure the next equality.
Finally, it is simple to check that both $b_1$ and $b_2$ are $O(1/n^{2s-2})$ if 
$c \leq (k-1)(1-1/s)$.
\end{proof}

We now provide the more general theorem.  

\begin{theorem} \label{thm:poissapproxext}
Consider an asymptotic regime where as $n \to \infty$, we have $k,s =
O(1)$ with $s \ge 2$, $\E[R^{2s}] = O(n^{-a})$ for some constant $2 <
a \le 2s$, and $t = O(n^c)$ for some positive constant $c$.  Let $U$
be a Poisson random variable such that
$\E[U]=\E[\hat{Q}_{k,s}]=\lambda<\infty$.  If
\[
c \le \frac{(k-1)(a-2) + \min(2a-6,0)}{2s},
\]
then the variation distance between the distributions ${\cal
L}(\hat{Q}_{k,s})$ of $\hat{Q}_{k,s}$ and
${\cal L}(U)$ of $U$ satisfies
\[
    \left\| {\cal L}(\hat{Q}_{k,s})-{\cal L}(U) \right\|
    = O(1/n).
\]
\end{theorem}
\begin{proof}
Applying Theorem~\ref{arratia} gives
\[
    \left\|{\cal L}(\hat{Q}_{k,s})-{\cal L}(U)\right\| \leq b_1 + b_2
\]
where
\[
    b_1  =  \sum_{X:|X|=k} \sum_{Y\in I(X)} p_X p_Y
\]
and
\[
    b_2 = \sum_{X:|X|=k} \sum_{Y\neq X\in I(X)} \E[Z_X Z_Y].
\]

We now evaluate $b_1$ and $b_2$.  Letting $\vec{R}$ denote the vector
of the $R_x$'s, we have that for any set $X$ of $k$ items
\[
    \Pr(Z_X = 1\ |\ \vec{R}) \le \binom{t}{s} \prod_{x \in X} R_x^s.
\]
Since the $R_x$'s are independent with common distribution $R$,
\[
    p_X = \E[\Pr(Z_X = 1\ |\ \vec{R})] \le \binom{t}{s} \E[R^s]^k.
\]

Using Jensen's inequality, we now have

\begin{align*}
    b_1  &=  \sum_{X:|X|=k} \sum_{Y\in I(X)} p_X p_Y\\
    &\le \left(\binom{n}{k}^2 - \binom{n}{k}\binom{n-k}{k}\right) \binom{t}{s}^2 \E[R^s]^{2k}\\
    &\le \binom{n}{k}^2\left(1 - \frac{\binom{n-k}{k}}{\binom{n}{k}}\right) \binom{t}{s}^2 \E[R^{2s}]^k\\
    &=\binom{n}{k}^2\left(1 - \prod_{i=0}^{k-1} \frac{n-k-i}{n-i}\right) \binom{t}{s}^2 \E[R^{2s}]^k\\
    &=\Theta(n^k)^2 \cdot \Theta(1/n) \cdot O(n^{2cs}) \cdot O(n^{-ka})\\
    &= O(n^{k(2-a) + 2cs - 1})\\
\end{align*}
We now turn our attention to $b_2$.  Consider sets $X \ne Y$ of $k$
items, and suppose $g = |X \cap Y|>0$.  If $Z_X
Z_Y = 1$, there exist disjoint subsets $A,B,C \in \{1,\ldots,t\}$
such that $0 \le |A| \le s$, $|B| = |C| = s-|A|$, all of the
transactions in $A$ contain both $X$ and $Y$, all of the transactions
in $B$ contain $X$, and all of the transactions in $C$ contain $Y$.
Therefore,
\begin{align*}
    \E[Z_X Z_Y\ |\ \vec{R}]
    &\le \sum_{i=0}^s \binom{t}{i; s-i; s-i} 
        \left(\prod_{x \in X \cup Y} R_x^i\right) \\
    & ~~~~~~~~~~~~~~~\times \left(\prod_{x \in X} R_x^{s-i}\right)
        \left(\prod_{y \in Y} R_y^{s-i}\right)\\
    &= \sum_{i=0}^s \binom{t}{i; s-i; s-i}
        \left(\prod_{x \in X \cap Y} R_x^{2s-i}\right) \\
    & ~~~~~~~~~~~~~~~\times \left(\prod_{x \in X - Y} R_x^s\right)
        \left(\prod_{y \in Y - X} R_y^s\right).
\end{align*}
Applying independence of the $R_x$'s and Jensen's inequality gives
\begin{align*}
    \E[Z_X Z_Y]
    &= \E[\E[Z_X Z_Y\ |\ \vec{R}]]\\
    &\le \sum_{i=0}^s \binom{t}{i; s-i; s-i} \E[R^{2s-i}]^g \E[R^s]^{2(k-g)}\\
    &\le \sum_{i=0}^s t^{2s-i} \E[R^{2s}]^{\frac{g(2s-i)}{2s}} \E[R^{2s}]^{k-g}\\
    &= \sum_{i=0}^s t^{2s-i} \E[R^{2s}]^{k - ig/2s}\\
    &\le O(1) \sum_{i=0}^s n^{(2s-i)c - a\left(k - ig/2s\right)}\\
    &= O(n^{2sc - ak}) \sum_{i=0}^s n^{i\left(\frac{ag}{2s} - c\right)}\\
   &= O\left(n^{2sc - ak + \max\left\{0, s\left(\frac{ag}{2s} - c\right)\right\} }\right) \\
\end{align*}
It follows that
\begin{align*}
    b_2 &\le \sum_{g=1}^{k-1} \binom{n}{g;k-g;k-g} O\left(n^{2sc - ak + \max\left\{0, s\left(\frac{ag}{2s} - c\right)\right\} }\right) \\
    &= O(n^{2k + 2sc - ak}) \sum_{g=1}^{k-1} n^{-g} O\left(n^{\max\left\{0, s\left(\frac{ag}{2s} - c\right)\right\} }\right) \\
\end{align*}
Now, for $2sc/a < g < k$, we have (using the fact that $a \ge 2$)
\begin{align*}
    n^{-g}n^{\max\left\{0, s\left(\frac{ag}{2s} - c\right)\right\}}
    = n^{g(\frac{a}{2} - 1)-sc} \le n^{(k-1)(\frac{a}{2} - 1)-sc}.
\end{align*}
Thus 
\[
b_2 = O(n^{2k + sc - ak + (k-1)(\frac{a}{2} - 1)}).
\]
(Here we
are using the fact that our choice of $c$ satisfies $c \le (k-1)(a -
2)/2s$ to ensure that $n^{(k-1)(\frac{a}{2} - 1) - cs} = \Omega(1)$.)

Now, we have
\[
b_1 = O(1/n)
\] 
since
\[ 
c \le \frac{(k-1)(a-2)}{2s} \le \frac{k(a-2)}{2s},
\]
and 
\[
b_2 = O(1/n) 
\]
since
\[ 
c \le \frac{k(a-2) + (a-4)}{2s}.
\]
Thus
\[
b_1 + b_2 =
O(1/n).
\]
\end{proof} 

It is easy to see that for fixed $k$, the quantities $b_1$ and $b_2$
defined in Theorem~\ref{arratia} are both decreasing in $s$.  In the
following, we will use the notation $b_1(s)$ and $b_2(s)$ to indicate
explicitly that both quantities are functions of $s$.  Therefore, for
a chosen $\epsilon$, with $0 < \epsilon < 1$, we can define
\begin{equation} \label{smin}
s_{\min} = \min \{s \geq 1 \; : \; b_1(s)+b_2(s) \leq \epsilon \}.
\end{equation}

It immediately follows that for every $s$ in the range $[s_{\min},\infty)$, 
the variation distance between the distribution of $\hat{Q}_{k,s}$ and
the distribution of a Poisson variable with the same expectation is
less than $\epsilon$. In other words, for every $s \geq s_{\min}$ the
number of $k$-itemsets with support at least $s$ is well approximated
by a Poisson variable. Theorems~\ref{thm:poissapproxsimple} and
\ref{thm:poissapproxext} proved above establish 
the existence of meaningful ranges of
$s$ for which the Poisson approximation holds, under certain
constraints on the individual item frequencies in the random dataset
and on the other parameters.

\subsection{A Monte Carlo method for determining $s_{\min}$}

While the analytical results of the previous subsection require that
the individual item frequencies in the random dataset be drawn from a
given distribution, in what follows we give experimental evidence
that the Poisson approximation for the distribution of $\hat{Q}_{k,s}$
holds also when the item frequencies are fixed arbitrarily, as is the
case of our reference random model. More specifically, we present a
method which approximates the support threshold $s_{\min}$ defined by
Equation~\ref{smin}, based on a simple Monte Carlo simulation which
returns estimates of $b_1(s)$ and $b_2(s)$. This approach is also
convenient in practice since it avoids the inevitable slack due to the
use of asymptotics in Theorem~\ref{thm:poissapproxext}.

For a given configuration of item frequencies and number of transactions,
let $\tilde{s}$ be the maximum expected support of any $k$-itemset
in a random dataset sampled according to that configuration,
that is, the product of the $k$ largest item frequencies. 
Conceivably, the value $b_1(\tilde{s})$ is rather large,
hence it makes sense to search for an $s_{\min}$ larger than $\tilde{s}$.
We generate $\Delta$ random datasets and
from each such dataset we mine all of the  $k$-itemsets of support at
least $\tilde{s}$. Let $W$ be the set of itemsets extracted in this
fashion from all of the generated datasets. For
each $s \geq \tilde{s}$ we can estimate $b_1(s)$ and $b_2(s)$ by
computing for each $X \in W$ the empirical probability $p_X$ of the
event $Z_X=1$, and for each pair $X,Y \in W$, with $X \cap Y \neq
\emptyset$, the empirical probability $p_{X,Y}$ of the event $(Z_X=1)
\wedge (Z_Y=1)$. Note that for itemsets not in $W$ these probabilities
are estimated as 0. If it turns out that
$b_1(\tilde{s})+b_2(\tilde{s}) > \epsilon/4$, then we let $\hat{s}_{\min}$
be the minimum $s > \tilde{s}$ such that $b_1(s)+b_2(s) \leq
\epsilon/4$. Otherwise, if $b_1(\tilde{s})+b_2(\tilde{s}) \leq
\epsilon/4$, we repeat the above procedure starting from $\tilde{s}/2$.
(Based on the above considerations this latter case will be unlikely.) 
Algorithm 1 implements the above ideas.

The following theorem provides 
a bound on the probability that $\hat{s}_{\min}$ be a conservative
estimate of $s_{\min}$, that is, $\hat{s}_{\min} \geq s_{\min}$.
\begin{theorem}
If $\Delta=O\left(\log(1/\delta)/\epsilon\right)$,
the output $\hat{s}_{\min}$ of the Monte-Carlo process satisfies
\[
\Pr(b_1(\hat{s}_{\min})+b_2(\hat{s}_{\min}) \leq \epsilon)\geq 1-\delta.
\]
\end{theorem}
\begin{proof}
Let assume $b_1(\hat{s}_{\min}) + b_2(\hat{s}_{\min}) > \epsilon$. Note that
$b_1(\hat{s}_{\min}) \leq b_2(\hat{s}_{\min})$, therefore we have
$b_2(\hat{s}_{\min}) > \epsilon/2$.
Let $B$ be the random variable corresponding to $\Delta$ times the estimate
of $b_2(\hat{s}_{\min})$ obtained with Algorithm
1. Thus $E[B] > \Delta \epsilon / 2$. Since Algorithm
1 returns $\hat{s}_{\min}$ as estimate of $s_{\min}$, we have that $B \le
\Delta \epsilon/4$.
Let 
\[
\Delta = \frac{8 \log(1/\delta)}{\epsilon},
\]
and $c < 1$ be such that:
\[
 (1-c)E[B] = \Delta \epsilon/4.
\]
Since $E[B] > \Delta \epsilon / 2$, we have $c \geq 1/2$.
Using Chernoff bound, we have that:
\begin{align*}
\Pr( B \le \Delta \epsilon/4 ) & \leq e^{- \frac{c^2 E[B]}{2}} \\
& \leq e^{- \frac{1}{4}\frac{ 8 \log(1/\delta)}{2}} \leq \delta.
\end{align*}
Thus $\Pr(b_1(\hat{s}_{\min}) + b_2(\hat{s}_{\min}) > \epsilon) \leq
\delta$.
\end{proof}

\begin{algorithm}[h]
\caption{FindPoissonThreshold}
\label{alg:montecarlo}
\begin{algorithmic}[1]
\REQUIRE Dataset $\D$ of $t$ transactions over $n$ items, 
vector $\vec{f}$ of item frequencies, $k$, $\Delta$, $\varepsilon$;
\ENSURE Estimate $\hat{s}_{\min}$ of $s_{\min}$;
\STATE $\tilde{s} \gets $ highest expected support of a $k$-itemset; 
\STATE $s_{\max} \gets 0$;
\STATE $W \gets \emptyset$;\label{line:alg}
\FOR {$i \gets 1$ to $\Delta$} \label{line2:alg}
\STATE $\hat{\D}_i \gets$ random dataset with parameters
$t$,$n$,$\vec{f}$;
\STATE $W \gets W \cup \left\{ \mbox{frequent }k\mbox{-itemsets in
}\hat{\D_i} \mbox{ w.r.t. }\tilde{s} \right\}$;
\ENDFOR
\IF {$W = \emptyset$}
\STATE $\tilde{s} \gets \tilde{s}/2$;
\STATE {\bf goto} \ref{line2:alg};
\ENDIF
\IF {($s_{\max} =0$)}
\STATE $s_{\max} \gets \displaystyle \max_{X \in W, \hat{\D}_i }\left\{
\mbox{support of }X \mbox{ in } \hat{\D}_i\right\} + 1$;
\ENDIF
\FOR{$s\gets \tilde{s}$ to $s_{\max}$}
\FORALL {$X \in W$}
\STATE $p_X(s) \gets $ empirical probability of \{$Z_X=1$\};
\ENDFOR
\FORALL {$X,Y \in W: X \cap Y \neq \emptyset$}
\STATE $p_{X,Y}(s) \gets$ empirical probability of \{$Z_{X,Y}=1$\};
\ENDFOR
\STATE $b_{1}(s) \gets \displaystyle \sum_{X,Y \in W; Y\in I(X)} p_X(s)
p_Y(s)$;
\STATE $b_{2}(s) \gets \displaystyle \sum_{X,Y \in W; X \neq Y\in I(X)}
p_{X,Y}(s)$;
\ENDFOR
\IF {$b_{1}(\tilde{s})+b_{2}(\tilde{s}) \leq \varepsilon/4$}
\STATE $s_{\max} \gets \tilde{s}$;
\STATE $\tilde{s} \gets \tilde{s}/2$; 
\STATE {\bf goto} \ref{line:alg};
\ENDIF
\STATE $\hat{s}_{\min} \gets \min\left\{s > \tilde{s}: b_{1}(s) + b_{2}(s)
\leq 
\varepsilon/4 \right\}$;
\STATE {\bf return} $\hat{s}_{\min}$;
\end{algorithmic}
\end{algorithm}
\setcounter{algorithm}{0}

For each dataset $\D$ of Table~\ref{tab:datasets} and for itemset
sizes $k=2,3,4$, we applied Algorithm 1 setting $\Delta = 1,000$ and
$\epsilon=0.01$. The values of $\hat{s}_{\min}$ we obtained are reported in
Table~\ref{fig:realpoissapprox} (we added the prefix ``Rand'' to each
dataset name, to denote the fact that the dataset is random and
features the same parameters as the corresponding real one).

\begin{table}[ht]
\begin{center}
\begin{tabular}{lccc}
\hline
 & \multicolumn{3}{c}{ $\hat{s}_{\min}$ }\\
\cline{2-4}
Dataset & $k = 2$ & $ k = 3$ & $ k = 4$\\
\hline
RandRetail & 9237 & 4366 & 784 \\
RandKosarak & 273266 & 100543 & 20120  \\
RandBms1 & 268 & 23 & 5 \\
RandBms2 & 168 & 13 & 4 \\
RandBmspos & 76672 & 15714 & 2717 \\
RandPumsb$^*$ & 29303 & 21893 & 16265 \\
\hline
\end{tabular}
\end{center}
\caption{\boldmath Values of $\hat{s}_{\min}$ for 
$\epsilon = 0.01$ and for $k=2,3,4$, in random datasets with the same
values of $n$, $t$, and with the same frequencies of the items as
the corresponding benchmark datasets.} \label{fig:realpoissapprox}
\end{table}

\section{Procedures for the discovery of high-support 
significant itemsets} \label{sec:methodology}

For a give itemset size $k$, the value $s_{\min}$ identifies a region
of (relatively high) supports where we concentrate our quest for
statistically significant $k$-itemsets.  In this section we develop
procedures to identify a family of $k$-itemsets (among those of
support greater than or equal to $s_{\min}$) which are statistically
significant with a controlled FDR.  More specifically, in
Subsection~\ref{sec:chernoff} we show that a family with the desired
properties can be obtained as a subset of the frequent $k$-itemsets
with respect to $s_{\min}$, selected based on a standard
multi-comparison test. However, the returned family may turn out to be
too small (i.e., the procedures yields a large number of false
negatives). To achieve higher effectiveness, in
Subsection~\ref{sec:sstar} we devise a more sophisticated procedure
which identifies a support threshold $s^* \geq s_{\min}$ such that
\emph{all} frequent $k$-itemsets with respect to $s^*$ are
statistically significant with a controlled FDR. In the next section
we will provide experimental evidence that in many cases the latter
procedure yields much fewer false negatives.

\subsection{A procedure based on a standard multi-comparison 
test}\label{sec:chernoff}

We present a first, simple procedure to discover significant itemsets with
controlled FDR, based on the following well established 
result in multi-comparison testing.
\begin{theorem}[\cite{BY01}]
\label{by-test}
Assume that we are testing for $m$ null hypotheses.  Let $p_{(1)}\leq
p_{(2)}\leq \dots\leq p_{(m)}$ be the ordered observed $p$-values of
the $m$ tests. For a given parameter $\beta$, with $0 < \beta < 1$, define
\begin{eqnarray}
\label{eqn:by}
\ell=\max\left\{i \geq 0: p_{(i)}\leq \frac{i}{m\sum_{j=1}^m \frac{1}{j}} \beta\right\},
\end{eqnarray}
and reject the null hypotheses corresponding to tests
$(1),\dots,(\ell)$.  Then, the FDR for the set of rejected null
hypotheses is upper bounded by $\beta$.
\end{theorem}

Let $\D$ denote an input dataset consisting of $t$ transactions over
$n$ items, and let $k$ be the fixed itemset size. Recall that $s_{\min}$
is the minimum support threshold for which the distribution of
$\hat{Q}_{k,s}$ is well approximated by a Poisson distribution.
First, we mine from $\D$ the set of frequent $k$-itemsets ${\cal
  F}_{(k)}(s_{\min})$.  Then, for each $X \in {\cal
  F}_{(k)}(s_{\min})$, we test the null hypothesis $H_0^X$ that the
observed support of $X$ in $\D$ is drawn from a Binomial distribution
with parameters $t$ and $f_X$ (the product of the individual
frequencies of the items of $X$), setting the rejection threshold as
specified by condition (\ref{eqn:by}), with parameters $\beta$ and $m
= {n \choose k}$.  Based on Theorem~\ref{by-test}, the itemsets of
${\cal F}_{(k)}(s_{\min})$ whose associated null hypothesis is
rejected can be returned as significant, with FDR upper bounded by
$\beta$.  The pseudocode Procedure~\ref{alg:fdrbino} implements the
strategy described above.

\floatname{algorithm}{Procedure}
\begin{algorithm}
\caption{}
\label{alg:fdrbino}
\begin{algorithmic}
\REQUIRE Dataset $\D$ of $t$ transactions over $n$ items, 
vector $\vec{f}$ of item frequencies, $k$, $\beta \in (0,1)$;
\ENSURE Family of significant $k$-itemsets with FDR $\leq \beta$;
\STATE Determine $s_{\min}$ and compute ${\cal F}_{(k)}(s_{\min})$ from $\D$;
\FORALL {$X \in {\cal F}_{(k)}(s_{\min})$}
	\STATE $s_X \gets $ support of $X$ in ${\cal D}$; 
	\STATE $f_X \gets \Pi_{i\in X}f_i$; 
	\STATE $p^{(X)} \gets \Pr(\mbox{Bin}(t , f_X) \geq s_X)$;
\ENDFOR
\STATE Let $p_{(1)}, p_{(2)}, \dots ,$ be the sorted sequence of the values 
$p^{(X)}$, with $X\in {\cal F}_{(k)}(s_{\min})$;
\STATE $m \gets {n \choose k}$;
\STATE $\ell = \max\left\{0, i: p_{(i)}\leq \frac{i}{m\sum_{j=1}^m \frac{1}{j}}\beta\right\}$;
\RETURN $\left\{ X \in {\cal F}_{(k)}(s_{\min}): p^{(X)} = p_{(i)}, 1\leq i\leq \ell \right\}$;
\end{algorithmic}
\end{algorithm}

\subsection{Establishing a support threshold for  
significant frequent itemsets} \label{sec:sstar}

Let $\alpha$ and $\beta$ be two constants in $(0,1)$.  We seek a
threshold $s^*$ such that, with confidence $1-\alpha$, the
$k$-itemsets in ${\cal F}_{(k)}(s^*)$ can be flagged as statistically
significant with FDR at most $\beta$.  The threshold $s^*$ is
determined through a robust statistical approach which ensures that
the number $Q_{k,s^*} = |{\cal F}_{(k)}(s^*)|$ deviates significantly
from what would be expected in a random dataset, and that the
magnitude of the deviation is sufficient to guarantee the bound on the
FDR.

Let $s_{\min}$ be the minimum support such that the Poisson
approximation for the distribution of $\hat{Q}_{k,s}$ holds for $s
\geq s_{\min}$, and let $s_{\max}$ be the maximum support of an item
(hence, of an itemset) in $\D$. Our procedure performs $h=\lfloor
\log_2 (s_{\max}-s_{\min}) \rfloor + 1$ comparisons. Let
$s_0=s_{\min}$ and $s_i =s_{\min}+2^i$, for $ 1 \leq i < h$.  In the
$i$-th comparison, with $0 \leq i < h$, we test the null hypothesis
$H_0^i$ that the observed value $Q_{k,s_i}$ is drawn from the same
Poisson distribution as $\hat{Q}_{k,s_i}$. We choose as $s^*$ the
minimum of the $s_i$'s, if any, for which the null hypothesis $H_0^i$
is rejected.

For the correctness of the above procedure, it is crucial to specify a
suitable rejection condition for each $H_0^i$.  Assume first that, for
$0 \leq i < h$, we reject the null hypothesis $H_0^i$ when the
$p$-value of the observed value $Q_{k,s_i}$ is smaller than
$\alpha_i$, where the $\alpha_i$'s are chosen so that
$\sum_{i=0}^{h-1} \alpha_i = \alpha$. Then, the union bound shows that
the probability of rejecting any true null hypothesis is less than
$\alpha$. However, this approach does not yield a bound on the FDR for
the set ${\cal F}_{(k)}(s^*)$.  In fact, some itemsets in ${\cal
  F}_{(k)}(s^*)$ are likely to occur with high support even under
$H_0^i$, hence they would represent false discoveries.  The impact of
this phenomenon can be contained by ensuring that the FDR is below a
specified level $\beta$. To this purpose, we must strengthen the
rejection condition, as explained below.

Fix suitable values $\beta_0, \beta_1, \ldots, \beta_{h-1}$ such that
$\sum_{i=0}^{h-1} \beta_i^{-1} \leq \beta$.  For $0 \leq i < h$, let
$\lambda_i = E[\hat{Q}_{k,s_i}]$. We now reject $H_0^i$ when the
$p$-value of $Q_{k,s_i}$ is smaller than $\alpha_i$, \emph{and}
$Q_{k,s_i} \geq \beta_i \lambda_i$.  The following theorem establishes
the correctness of this approach.
\begin{theorem} \label{correctness-p2}
With confidence $1-\alpha$, ${\cal F}_{(k)}(s^*)$ 
is a family of statistically significant frequent
$k$-itemsets with FDR at most $\beta$.
\end{theorem}
\begin{proof}
Observe that since $\sum_{i=0}^{h-1} \alpha_i \leq \alpha$, we have
that all rejections are correct, with probability at least
$1-\alpha$. Let $E_i$ be the event \textit{``$H^i_0$ is rejected''} or
equivalently, \textit{``the $p$-value of $Q_{k,s_i}$ is smaller than
  $\alpha_i$ and $Q_{k,s_i} \geq \beta_i \lambda_i$''}.  Suppose that
$H^i_0$ is the first rejected null hypothesis, for some index $i$,
whence $s^*= s_i$. In this case, $Q_{k,s_i}$ itemsets are flagged as
significant.  We denote by $V_i$ the number of false discoveries among
these $Q_{k,s_i}$ itemsets.  It is easy to argue that the expectation
of $V_i$ is upper bounded by $E[X_i | E_i,
  \bar{E}_{i-1},\dots,\bar{E}_0]$, where $X_i$ is a Poisson variable
with expectation $\lambda_i$. Since $Q_{k,s_i} \geq \beta_i \lambda_i$
when $H^i_0$ is rejected, by the law of total probability we have
\begin{eqnarray*}
FDR & \leq & \sum_{i=0}^{h-1} E\left[\frac{V_i}{Q_{k,s_i}}\right]
\Pr(E_i,\bar{E}_{i-1},\dots,\bar{E}_0) \\
& \leq & \sum_{i=0}^{h-1} \frac{E\left[V_i\right]}{\beta_i \lambda_i} 
\Pr(E_i,\bar{E}_{i-1},\dots,\bar{E}_0) \\
& \leq & \sum_{i=0}^{h-1} 
         \frac{E[X_i~|~E_i \bar{E}_{i-1},\dots,\bar{E}_0]}{\beta_i \lambda_i} 
         \Pr(E_i,\bar{E}_{i-1},\dots,\bar{E}_0) \\
& = & \sum_{i=0}^{h-1} 
         \frac{\sum_{j\geq 0}j \Pr(X_i=j,E_i,\bar{E}_{i-1},\dots,\bar{E}_0)}
              {\beta_i \lambda_i} \\
& \leq & \sum_{i=0}^{h-1} \frac{\lambda_i}{\beta_i \lambda_i} =
\sum_{i=0}^{h-1} \frac{1}{\beta_i} \leq \beta. 
\end{eqnarray*}
\end{proof}

The pseudocode Procedure~\ref{alg:test1} specifies more formally our approach
to determine the support threshold $s^*$. Note that estimates for the
$\lambda_i$'s needed in the for-loop of Lines 7-9 can be obtained from
the same random datasets generated in Algorithm~\ref{alg:montecarlo},
which are used there for the estimation of $s_{\min}$.

\begin{algorithm}
\caption{}
\label{alg:test1}
\begin{algorithmic}[1]
\REQUIRE Dataset $\D$ of $t$ transactions over $n$ items, 
vector $\vec{f}$ of item frequencies, $k$, $\alpha, \beta \in (0,1)$;
\ENSURE $s^*$  such that, with confidence $1-\alpha$,
${\cal  F}_{(k)}(s^*)$ is a family of significant $k$-itemsets
with FDR $\leq \beta$;
\STATE Determine $s_{\min}$ and compute ${\cal F}_{(k)}(s_{\min})$ from $\D$;
\STATE $s_{\max}\gets$ maximum support of an item;
\STATE $i \gets 0$;  $s_0\gets s_{\min}$; 
\STATE $h \gets \lfloor \log_2 (s_{\max}-s_{\min}) \rfloor+1$;
\STATE Fix $\alpha_0,\dots,\alpha_{h-1} \in (0,1)$ s.t.
$\sum_{i=0}^{h-1}\alpha_i = \alpha$;
\STATE Fix $\beta_0,\dots,\beta_{h-1} \in (0,1)$ s.t. 
$\sum_{i=0}^{h-1}\beta_i^{-1} = \beta$;
\FOR{$i \gets 0$ to $h-1$}
\STATE Compute $\lambda_i = E[\hat{Q}_{k,s_i}]$;
\ENDFOR
\WHILE{ $i < h$ }
\STATE Compute $Q_{k,s_i}$;
\IF {$(\Pr(\mbox{Poisson}(\lambda_i) \geq Q_{k,s_i}) \leq \alpha_i)$ 
    {\bf and} ($Q_{k,s_i}\geq \beta_i \lambda_i$)}
\RETURN $s^* \gets s_i$;
\ENDIF
\STATE	$s_{i+1} \gets s_{\min} + 2^{i+1} $; 
\STATE $i \gets i+1$;
\ENDWHILE
\RETURN $s^* \gets \infty$;
\end{algorithmic}
\end{algorithm}

\section{Experimental Results} \label{sec:experiments}

In order to show the potential of our approach, in this section we
report on a number of experiments performed on the benchmark datasets
of Table~\ref{tab:datasets}.  First, in Subsection~\ref{sec:realdata},
we validate experimentally the methodology implemented by
Procedure~\ref{alg:test1}, while in Subsection~\ref{sec:chernoffresults}, we
compare Procedure~\ref{alg:test1} against the more standard
Procedure~\ref{alg:fdrbino}, with respect to their ability to
discover significant itemsets.

\subsection{Experiments on benchmark datasets} \label{sec:realdata}

For each benchmark dataset in Table~\ref{tab:datasets} and for
$k=2,3,4$, we apply Procedure~\ref{alg:test1} with
$\alpha=\beta=0.05$, and $\alpha_i=\beta_{i}^{-1}=0.05/h$. The results
are displayed in Table~\ref{res:test3}, where, for each dataset and
for each value of $k$, we show: the support $s^*$ returned by
Procedure~\ref{alg:test1}, the number $Q_{k,s^*}$ of $k$-itemsets with
support at least $s^*$, and the expected number $\lambda(s^*)$ of
itemsets with support at least $s^*$ in a corresponding random
dataset.

\begin{table*}[ht]
\begin{center}
{\footnotesize
\begin{tabular}{lccccccccccc}
\hline
 & \multicolumn{3}{c}{$k=2$} & &\multicolumn{3}{c}{$k=3$} 
 & & \multicolumn{3}{c}{$k=4$} \\
\cline{2-4} \cline{6-8} \cline{10-12}
Dataset & $s^*$ & $Q_{k,s^*}$ & $\lambda(s^*)$ & & $s^*$ & $Q_{k,s^*}$ &
$\lambda(s^*)$ & & $s^*$ & $Q_{k,s^*}$ & $\lambda(s^*)$  \\
\hline
Retail & $\infty$ & 0 & 0 & & $\infty$ & 0 & 0 & & 848 & 6 & 0.01 \\
Kosarak & $\infty$ & 0 & 0 & & $\infty$ & 0  & 0 & & 21144 & 12 & 0.01 \\
Bms1 & 276  & 56 & 0.19 & & 23 & 258859 & 0.06 & & 5 & ~27M & 0.05 \\
Bms2 & 168 & 429 & 0.73 & & 13 & 36112 & 0.25 & & 4 & 714045 & 0.01 \\
Bmspos & $\infty$ & 0 & 0 & & 16226 & 22 & 0.01 & & 2717 & 891 & 0.38 \\
Pumsb$^*$ & 29303 & 29 & 0.05 & & 21893 & 406 & 0.35 & & 16265 & 6293 & 1.37
\\
\hline
\end{tabular}
}
\end{center}
\caption{Results obtained by applying Procedure~\ref{alg:test1} with 
$\alpha = 0.05, \beta = 0.05$ and $k= 2,3,4$ to the benchmark 
datasets of Table~\ref{tab:datasets}.}\label{res:test3}
\end{table*}

We observe that for most pairs (dataset,$k$) the number of significant
frequent $k$-itemsets obtained is rather small, but, in fact, at
support $s^*$ in random instances of those datasets, less than two
(often much less than one) frequent $k$-itemsets would be expected.
These results provide evidence that our methodology not only defines
significance on statistically rigorous grounds, but also provides the
mining task with suitable support thresholds that avoid explosion of
the output size (the widely recognized ``Achilles' heel'' of
traditional frequent itemset mining). This feature crucially relies on
the identification of a region of ``rare events'' provided by the
Poisson approximation.  As discussed in Section~\ref{sec:ourresults},
the discovery of significant itemsets with low support (not returned
by our method) would require the extraction of a large (possibly
exponential) number of itemsets, that would make any strategy aiming
to discover these itemsets unfeasible. Instead, we provide an efficient
method to identify, with high confidence level, the family of most
frequent itemsets that are statistically significant without
overwhelming the user with a huge number of discoveries.

There are, however, a few cases where the number of itemsets returned
is still considerably high. Their large number may serve as a sign
that the results call for further analysis, possibly using clustering
techniques \cite{XinHYC05} or limiting the search to \emph{closed
  itemsets} \cite{PasquierBTL99}\footnote{An itemset is \emph{closed}
  if it is not properly contained in another itemset with the same
  support.}. For example, consider dataset Bms1 with $k=4$ and the
corresponding value $s^{*} = 5$ from Table~\ref{res:test3}. Extracting
the closed itemsets of support greater or equal to $s^{*}$ in that
dataset revealed the presence of a closed itemset of cardinality $154$
with support greater than $7$ in the dataset. This itemset, whose
occurrence by itself represents an extremely unlikely event in a
random dataset, accounts for more than 22M non-closed subsets with the
same support among the 27M reported as significant.

It is interesting to observe that the results obtained for dataset
Retail provide further evidence for the conclusions drawn in
\cite{GionisMMT06}, which suggested random behavior for this dataset
(although the random model in that work is slightly different from
ours, in that the family of random datasets also maintains the same
transaction lengths as the real one).  Indeed, no support threshold
$s^*$ could be established for mining significant $k$-itemsets with $k
= 2,3$, while the support threshold $s^*$ identified for $k=4$ yielded
as few as 6 itemsets. However, the conclusion drawn in \cite{GionisMMT06}
was based on a qualitative assessment of the discrepancy between the
numbers of frequent itemsets in the random and real datasets, while
our methodology confirms the findings on a statistically sound
and rigorous basis.

Observe also that for some other pairs (dataset,$k$) our procedure
does not identify any support threshold useful for mining
statistically significant itemsets. This is an evidence that, for the
specific $k$ and for the high supports considered by our approach, these
datasets do not present a significant deviation from the corresponding
random datasets.

Finally, in order to assess its robustness, we applied our methodology
to random datasets. Specifically, for each benchmark dataset of
Table~\ref{tab:datasets} and for $k=2,3,4$, we generated 100 random
instances with the same parameters as those of the benchmark, and
applied Procedure~\ref{alg:test1} to each instance, searching for a
support threshold $s^*$ for mining significant itemsets.  In
Table~\ref{tab:randomdatatest3} we report the number of times
Procedure~\ref{alg:test1} was successful in returning a finite value
for $s^*$. As expected, the procedure returned $s^*=\infty$, in
\emph{all cases} but for 2 of the 100 instances of the random dataset
with the same parameters as dataset Pumsb$^*$ with $k=2$.  However, in
these two latter cases, mining at the identified support threshold only
yielded a very small number of significant itemsets (one and two,
respectively).

\begin{table}[h]
\begin{center}
\begin{tabular}{lccc}
\hline
& \multicolumn{3}{c}{$s^* < \infty$} \\
\cline{2-4}
Dataset & $k=2$ & $k=3$ & $k=4$ \\
\hline
RandomRetail & 0 & 0 & 0 \\
RandomKosarak & 0 & 0 & 0  \\
RandomBms1 & 0 & 0 & 0 \\
RandomBms2 & 0 & 0 & 0 \\
RandomBmspos & 0 & 0 & 0  \\
RandomPumsb$^*$ & 2 & 0 & 0 \\
\hline
\end{tabular}
\end{center}
\caption{Results for Procedure~\ref{alg:test1} with $\alpha = 0.05, \beta =
0.05$ for random versions of benchmark datasets; each entry
reports the number of times, out of 100 trials, the procedure
returned a finite value for $s^*$.}\label{tab:randomdatatest3}
\end{table}

\subsection{Relative effectiveness of Procedures~\ref{alg:fdrbino} 
and \ref{alg:test1}} \label{sec:chernoffresults}

In order to assess the relative effectiveness of the two procedures
presented in the previous section, we applied them to the benchmark
datasets of Table~\ref{tab:datasets}.  Specifically, we compared the number of
itemsets extracted using the threshold $s^*$ provided by
Procedure~\ref{alg:test1}, with the number of itemsets flagged as
significant using the more standard method based on Benjamini and
Yekutieli's technique (Procedure~\ref{alg:fdrbino}), imposing the same
upper bound $\beta=0.05$ on the FDR.

The results are displayed in Table~\ref{res:chernoffFDR}, where for
each pair (dataset,$k$), we report the cardinality of the family
${\cal R}$ of $k$-itemsets flagged as significant by
Procedure~\ref{alg:fdrbino}, and the ratio $r=Q_{k,s^*}/|{\cal R}|$,
where $Q_{k,s^*}$ is the number of $k$-itemsets of support at least
$s^*$, which are returned as significant with the methodology of
Subsection~\ref{sec:sstar}.

We observe that in all cases where Procedure~\ref{alg:test1} returned
a finite value of $s^*$ the ratio $r$ is greater than or equal to 1
(except for dataset Bms1 and $k=2$, and dataset Bmspos and $k=3$, where $r$
is however very close to 1).  Moreover, in some cases the ratio $r$ is
rather large.  Since both methodologies identify significant $k$-itemsets
among all those of support at least $s_{\min}$, these results provide
evidence that the methodology of Subsection~\ref{sec:sstar} is often more
(sometimes much more) effective.  The methodology succeeds in identifying
more significant itemsets, since it evaluates the significance of the
\emph{entire} set ${\cal F}_{(k)}(s^*)$ by comparing $Q_{k,s^*}$ to
$\hat{Q}_{k,s^*}$. In contrast, Procedure~\ref{alg:fdrbino} must
implicitly test considerably more hypotheses (corresponding to the
significance all possible $k$-itemsets), thus the power of the test
(1-$Pr$(Type-II error)) is significantly smaller.

Observe that the cases where $r=0$ in Table~\ref{res:chernoffFDR}
correspond to pairs (dataset,$k$) for which Procedure~\ref{alg:test1}
returned $s^*=\infty$, that is, the procedure was not able to identify
a threshold for mining significant $k$-itemsets. Note, however, that
in all of these cases the number of significant $k$-itemsets returned
by Procedure~\ref{alg:fdrbino} is extremely small (between 1 and 3).
Hence, for these pairs, both methodologies indicate that there is very
little significant information to be mined at high supports.

\begin{table}[h]
\begin{center}
\begin{tabular}{lcccccccc}
\hline
 & \multicolumn{2}{c}{$k=2$} & & \multicolumn{2}{c}{$k=3$} & &
\multicolumn{2}{c}{$k=4$} \\
\cline{2-3} \cline{5-6} \cline{8-9}
Dataset & $|{\cal R}|$ & $r$ & & $|{\cal R}|$ & $r$ & & $|{\cal R}|$ & $r$
\\
\hline
Retail & 3 & 0 & & 3 & 0 & & 6 & 1.0  \\
Kosarak & 1 & 0 & & 1 & 0 & & 12 & 1.0  \\
Bms1 & 60 & 0.933 & & 64367 & 4.441 & & 219706 & 122.9 \\
Bms2 & 429 & 1.0 & & 25906 & 1.394 & & 60927 & 11.72 \\
Bmspos & 2 & 0 & & 23 & 0.957 & & 891 & 1.0  \\
Pumsb$^*$ & 29  & 1.0 & & 406 & 1.0 & & 6288 & 1.001 \\
\hline
\end{tabular}
\end{center}
\caption{Results using Test~\ref{alg:fdrbino} to bound the FDR with $\beta =
0.05$ for itemsets of support $\geq s_{\min}$. }\label{res:chernoffFDR}
\end{table}

\section{Conclusions}\label{sec:conclusions} 

The main technical contribution of this work is the proof that in a random
dataset where items are placed independently in transactions, there is
a minimum support $s_{\min}$ such that the number of $k$-itemsets with
support at least $s_{\min}$ is well approximated by a Poisson
distribution. The expectation of the Poisson distribution and the
threshold $s_{\min}$ are functions of the number of transactions,
number of items, and frequencies of individual items.

This result is at the base of a novel methodology for mining frequent
itemsets which can be flagged as statistically significant incurring a
small FDR. In particular, we use the Poisson distribution as the
distribution of the null hypothesis in a novel multi-hypothesis
statistical approach for identifying a suitable support threshold $s^*
\geq s_{\min}$ for the mining task. We control the FDR of the output
in a way which takes into account global characteristics of the
dataset, hence it turns out to be more powerful than other standard
statistical tools (e.g., \cite{BY01}).  The results of a number of
experiments, reported in the paper, provide evidence of the
effectiveness of our approach.

To the best of our knowledge, our
methodology represents the first attempt at establishing a support
threshold for the classical frequent itemset mining problem with a
quantitative guarantee on the significance of the output.

\bibliographystyle{plain}

\end{document}